\documentclass[useAMS,usenatbib,onecolumn]{mn2e}

\usepackage{amssymb}
\usepackage{amsmath}
\usepackage{graphics}
\usepackage{graphicx}
\usepackage{epsfig,epstopdf,color,times}

\newcommand{\be}{\begin{equation}}
\newcommand{\ee}{\end{equation}}
\newcommand{\bea}{\begin{eqnarray}}
\newcommand{\eea}{\end{eqnarray}}

\title[Cosmological non-linear power spectra]{Cosmological nonlinear density and velocity power spectra including nonlinear vector and tensor modes}
\author[J. Hwang, D. Jeong and H. Noh]{Jai-chan Hwang$^{1}$\thanks{E-mail:
jchan@knu.ac.kr (JH); djeong@psu.edu (DJ); hr@kasi.re.kr (HN)}, Donghui Jeong$^{2,3}$\footnotemark[1], Hyerim Noh$^{4}$\footnotemark[1]\\
             ${}^{1}$Department of Astronomy and Atmospheric Sciences,
             Kyungpook National University, Daegu, 702-701, Korea \\
             ${}^{2}$Department of Astronomy and Astrophysics,
             The Pennsylvania State University, University Park, PA 16802, USA\\
             ${}^{3}$Institute for Gravitation and the Cosmos, The Pennsylvania State University, University Park, PA 16802, USA \\
             ${}^{4}$Korea Astronomy and Space Science Institute, Daejeon, 305-348, Korea}

\begin{document}
\date{}
\pagerange{\pageref{firstpage}--\pageref{lastpage}} \pubyear{2015}
\maketitle
\label{firstpage}

\begin{abstract}

We present the leading order non-linear density and velocity power spectra in the complete form; previous studies have omitted the vector- and tensor-type perturbations simultaneously excited by the scalar-type perturbation in nonlinear order. These additional contributions are comparable to the scalar-type purely relativistic perturbations, and thus negligible in the current paradigm of concordance cosmology: i.e., concerning density and velocity perturbations of the pressureless matter in perturbation regime well inside of matter-dominated epoch, we show that pure Einstein's gravity contributions appearing from the third order are entirely negligible (five orders of magnitude smaller than the Newtonian contributions) in all scales. We thus prove that Newtonian perturbation theory is quite reliable in calculating the amplitude of matter fluctuations even in the precision era of cosmology. Therefore, besides the ones imprinted as the initial condition generated in the earlier phase, the other relativistic effect relevant for interpreting observational data must be the projection effect that occurs when mapping galaxies on to the observed coordinate.
\end{abstract}


\section{Introduction}

Perturbation theory is a common practice in wide variety of physics problems. While nearly all fundamental equations in theoretical physics
are non-linear, in most cases only the linear perturbation can be
mathematically handled with sufficient confidence and precision.
Thus, it is truly a surprise to discover lately that many of the cosmologically relevant observations can be accommodated in a simplest geometric background world model with additional linear perturbations. These are temperature anisotropies and polarizations in cosmic microwave background radiation and the distribution of galaxies in the large scale (large-scale structure) in the observational side, and the Friedmann world model with small perturbations in the theoretical side \citep{Friedmann-1922,Lifshitz-1946}, now encapsulated in the concordance or precision cosmology. Although their large-scale effects are supposed to be small in the concordance model, in this precision era of cosmology, the non-linear perturbations might be important so that we need to include higher order corrections. In fact, the fully nonlinear situation appears as we pay attention to the smaller scales in the large-scale structure.

In cosmology, the linear perturbation is often decomposed into three types of perturbation. These are scalar-, vector- and tensor-types of perturbation \citep{Lifshitz-1946,York-1973,Bardeen-1980,Kodama-Sasaki-1984} where the last one is absent in the Newtonian approach. Such a three-type decomposition, however, is mathematically not unique to the non-linear order \citep{Hwang-Noh-2013}. Furthermore, as all three types
of perturbations are convoluted in the equation level, the physical meaning of each type naturally becomes obscured as well. Only to the linear order and in the spatially homogeneous and isotropic background, the three types of perturbations decouple in the equation level \citep{Lifshitz-1946}. Thus, only to the linear-order perturbation in the Friedmann background world model the scalar-, vector- and tensor-type perturbations can be properly interpreted as corresponding to the density, the rotation, and the gravitational wave perturbations, respectively. Even when we consider pure scalar-type perturbations to linear order, the vector- and tensor-type perturbations are naturally excited at non-linear order sourced by the linear scalar-type perturbation.

The leading non-linear power spectra demands perturbation expanded to the third order \citep{Vishniac-1983}. In our previous work, we presented the leading nonlinear power spectra of the density and velocity in the zero-pressure medium \citep{JGNH-2011}. We considered Einstein's gravity in the comoving gauge. In that work, we considered purely scalar-type perturbation, thus omitting the naturally accompanying effects of the vector- and tensor-type perturbations excited by the scalar-type one to the nonlinear order.

Here, we include these effects which start appearing from the third-order in perturbation. We shall show that the second-order vector- and tensor-perturbations generated from the linear scalar-perturbations contribute to the third-order solutions of density and velocity perturbations. We derive the general relativistic energy and momentum conservation equations including these effects, see equations (\ref{delta-eq}) and (\ref{u-eq}). In these equations, ${\bf Y}$ and $Z_{ij}$ terms are the contributions from the vector- and tensor-perturbations, respectively. We present the complete leading non-linear power spectra of the density and velocity in Figs \ref{PS-density} and \ref{PS-velocity}, respectively.

\section{Third-order perturbations in the comoving gauge}

We consider zero-pressure fluid in a flat background, thus set
$\widetilde p \equiv 0$ and $\overline K \equiv 0$ (variables are defined in Appendix \ref{sec:FNL-eqs}). Our aim is to derive third-order perturbation equations for the density and velocity perturbations in the comoving gauge without linear-order vector and tensor perturbations. Thus, in what follows, we consider only scalar-type perturbations in linear order and vector- and tensor-type perturbations shall be treated as higher order quantities. We decompose the perturbation variables to the scalar and vector types as \bea
   & & \widehat v_i \equiv - \widehat v_{,i} + \widehat v_i^{(v)}, \quad
       \chi_i \equiv c \chi_{,i} + \chi_i^{(v)},
\eea
with $\widehat v_i^{(v),i} \equiv 0 \equiv \chi_i^{(v),i}$; for our metric and energy-momentum convention, see Appendix \ref{sec:FNL-eqs}. By considering only scalar-type perturbations in the linear order, we set
$\widehat v_i^{(v)} \equiv 0 \equiv \chi_i^{(v)}$ to the linear order.
From the second order, the vector-type perturbations are generated from the scalar perturbations.

In our fully non-linear formulation \citep{Hwang-Noh-2013}, we have ignored the tensor-type perturbations. Just like the vector-type perturbations are generated from the scalar perturbations in nonlinear order, the tensor-type perturbations must be generated as well. The tensor perturbation generated from the scalar ones is at least second order in perturbation. Thus, in our calculation, aiming for third-order solutions for density and velocity perturbations, we only need to keep the tensor-type perturbations appearing with the scalar-type perturbation: we will recover this contribution in equation
(\ref{Raychaudhury-eq-correction}) below. Note that this is not the case for the vector perturbation as we shall show in what follows.

By taking the comoving gauge, we set
\bea
   & & \widehat v \equiv 0,
\eea
to all orders in perturbation.

We have identified $\delta$ and $\kappa$ as the density and velocity
perturbation variables, see equation (\ref{kappa-identify}); we have
$\widetilde \mu \equiv \widetilde \varrho c^2$, $\widetilde \varrho \equiv \varrho + \delta \varrho$ and $\delta \varrho \equiv \varrho \delta$. Equations for $\delta$ and $\kappa$ follow from equations (\ref{eq6}) and (\ref{eq4}), respectively. We note that the vector-type perturbations appear in the $\widehat v_i$ and $\chi_i$, and in equations (\ref{eq6}) and (\ref{eq4}), these terms appear at least in quadratic combination. As the vector perturbations generated from the scalar ones are at least second order, we need $\widehat v_i$ and the vector part of $\chi_i$ only to the second order. These can be read from equations (\ref{eq7}) and (\ref{eq3}), respectively.

For $\widehat v_i$, from equation (\ref{eq7}), we have
\bea
   & & \dot {\widehat v}_i^{(v)} + {\dot a \over a} \widehat v_i^{(v)}
       + {c^2 \over a} \delta {\cal N}_{,i}
       = {\rm non-linear \; terms \; all \; involving \;} \widehat v_i^{(v)},
\label{eq:vi_vec}
\eea
to the fully non-linear order. The equation (\ref{eq:vi_vec}) to the linear order reduces to $\delta {\cal N} = 0$. To the second order, terms in the right-hand side vanish, thus we have $\delta {\cal N} = 0$ and $\widehat v_i^{(v)} \propto a^{-1}$. As $\widehat v_i^{(v)}$ is purely decaying proportional to the reciprocal of the scale factor in an expanding medium, we may set $\widehat v_i^{(v)} = 0$ to the second order. This procedure can be continued perturbatively to all higher order perturbations, and we have
\bea
   & & \delta {\cal N} = 0, \quad
       \widehat v_i^{(v)} = 0, \quad
       {\rm thus} \quad
       \widehat v_i = 0,
\eea
to the fully non-linear order. The vector part of $\chi_i$ will be handled below.

Equations (\ref{eq6}) and (\ref{eq4}) to the third order, respectively,
\bea
   & & \dot \delta - \kappa
       + {c \over a^2} \left( 1 - 2 \varphi \right)
       \delta_{,i} \chi^i
       - \delta \kappa
       = 0,
   \label{energy-conservation-correction} \\
   & & \dot \kappa
       + 2 H \kappa
       - 4 \pi G \varrho \delta
       + {c \over a^2} \left( 1 - 2 \varphi \right) \kappa_{,i} \chi^i
       - {1 \over 3} \kappa^2
       - {c^2 \over a^4} \left( 1 - 4 \varphi \right)
       \left[ {1 \over 2} \chi^{i,j} \left( \chi_{i,j} + \chi_{j,i} \right)
       - {c^2 \over 3} \left( \Delta \chi \right)^2 \right]
   \nonumber \\
   & & \qquad
       + {4 c^4 \over a^4} \left(
       \chi^{,i} \varphi^{,j} \chi_{,ij}
       - {1 \over 3} \chi^{,i} \varphi_{,i} \Delta \chi \right)
       = {2 c^2 \over a^2} \chi^{,ij} \dot h_{ij},
   \label{Raychaudhury-eq-correction}
\eea
where $h_{ij}$ is the tensor-type perturbation generated from the scalar-type perturbation; see equation 21 in \cite{Hwang-Noh-2005-third} and equation 126 in \cite{Hwang-Noh-2007-Newtonian}; these works include full third order version of these equations including the vector- and tensor-type perturbations in the context of multiple zero-pressure fluids. In equation (\ref{Raychaudhury-eq-correction}), we kept only the second-order tensor-type perturbation generated from the scalar perturbation; this is the only surviving tensor perturbation term in our situation ignoring the linear tensor mode. In order to determine $\chi_i$, we need equation (\ref{eq3}) to the second order as
\bea
   & & \left( \kappa + c^2 {\Delta \over a^2} \chi \right)_{,i}
       + {3 \over 4} c {\Delta \over a^2} \chi_i^{(v)}
       = {c^2 \over a^2} \left[
       \left( 2 \varphi \Delta \chi
       - \varphi_{,j} \chi^{,j} \right)_{,i}
       + {3 \over 2} \left(
       \varphi_{,ij} \chi^{,j}
       + \chi_{,i} \Delta \varphi \right)
       \right],
   \label{mom-constraint-correction}
\eea
which can be further decomposed as following with scalar and vector
contributions $X$ and $Y_i$:
\bea
   & & c {\Delta \over a^2} \chi_i
       = - \kappa_{,i} + {1 \over a} X_{,i} + {1 \over a} Y_i,
   \nonumber \\
   & & \kappa + c^2 {\Delta \over a^2} \chi
       = {c^2 \over a^2} \left[ 2 \varphi \Delta \chi
       - \varphi_{,i} \chi^{,i}
       + {3 \over 2} \Delta^{-1} \nabla^i \left(
       \varphi_{,ij} \chi^{,j}
       + \chi_{,i} \Delta \varphi \right) \right]
       \equiv {1 \over a} X,
   \nonumber \\
   & & c {\Delta \over a^2} \chi_i^{(v)}
       = {2 c^2 \over a^2} \left[
       \varphi_{,ij} \chi^{,j}
       + \Delta \varphi \chi_{,i}
       - \nabla_i \Delta^{-1} \nabla^j \left(
       \varphi_{,jk} \chi^{,k}
       + \chi_{,j} \Delta \varphi \right)
       \right]
       \equiv {1 \over a} Y_i.
   \label{chi}
\eea %
In order to close the system of equations above, we need $\varphi$ to the linear order. From equation (\ref{eq2}) to the linear order, we have (with $\overline K = 0$)
\bea
   & & 4 \pi G \delta \varrho
       + {\dot a \over a} \kappa
       + c^2 {\Delta \over a^2} \varphi = 0.
   \label{varphi}
\eea

\section{General relativistic energy and momentum conservation equations}

By identifying the density and velocity perturbations 
\bea
   & & \delta \equiv {\delta \varrho \over \varrho}, \quad
       \kappa \equiv - {1 \over a} \nabla \cdot {\bf u}
       \equiv - {\Delta \over a} u,
   \label{kappa-identify}
\eea
we can reveal Newtonian correspondence of our equations derived in Einstein's gravity.

We find $\chi_i$ to second order from equation (\ref{chi}) as
\bea
   & & {c \over a} \vec{\chi}
       \equiv {c \over a} \left( c \nabla \chi
       + \vec{\chi}^{(v)} \right)
       = {\bf u}
       + \Delta^{-1} \left( \nabla X + {\bf Y} \right),
   \nonumber \\
   & & X
       \equiv 2 \varphi \nabla \cdot {\bf u}
       - {\bf u} \cdot \nabla \varphi
       + {3 \over 2} \Delta^{-1} \nabla \cdot \left(
       {\bf u} \cdot \nabla \nabla \varphi
       + {\bf u} \Delta \varphi \right),
   \nonumber \\
   & & {\bf Y}
       \equiv 2 \left[ {\bf u} \cdot \nabla \nabla \varphi
       + {\bf u} \Delta \varphi
       - \nabla \Delta^{-1} \nabla \cdot \left(
       {\bf u} \cdot \nabla \nabla \varphi
       + {\bf u} \Delta \varphi \right) \right].
   \label{XY}
\eea
The equations above are valid to the second order with divergence-free vector field ${\bf Y}$ (that is, $\nabla \cdot {\bf Y} = 0$). The linear-order equation reads $c^2 \chi = a u$.

Since $\varphi$ appears only with  second-order quantities in equations
(\ref{energy-conservation-correction}), (\ref{Raychaudhury-eq-correction}) and (\ref{XY}), we only need $\varphi$ to linear order. From equation (\ref{varphi}), we find
\bea
   & & c^2 {\Delta \over a^2} \varphi
       = - 4 \pi G \varrho \delta
       + {\dot a \over a} {1 \over a} \nabla \cdot {\bf u}.
   \label{varphi-eq2}
\eea
For $\overline K = 0$ (but with general $\Lambda$), we have $\varphi$ constant in time; see equation (\ref{varphi-linear-sol}) for the general solution. In particular for Einstein-de Sitter models ($\overline K = 0 = \Lambda$), we have
\bea
   & & \varphi (\equiv \varphi_v) = {5 \over 3} \varphi_\chi,
\eea
where $\varphi_v$ and $\varphi_\chi$ are the $\varphi$ (diagonal part of $\delta g_{ij}$) variable in the comoving gauge ($\widehat v \equiv 0$) and in the zero-shear gauge ($\chi \equiv 0$), respectively. We have (for general $\overline K$ and $\Lambda$ but without anisotropic stress)
\bea
   & & \varphi_\chi = - \alpha_\chi = {1 \over c^2} U,
\eea
where $U$ is perturbed Newtonian gravitational potential ($\alpha_\chi$ is $- \delta g_{00}$ part in the zero-shear gauge). The metric perturbation variable $\varphi$ (without taking the gauge condition) completely determines the intrinsic curvature perturbation of our fully non-linearly perturbed cosmological space-time, see equation 63 in Noh (2014).

While the vector contribution is accounted by the presence of $Y_i$ term via $\chi_i^{(v)}$ in equation (\ref{chi}), the tensor contribution is
accounted by $\dot h_{ij}$ in equation (\ref{Raychaudhury-eq-correction}). The tensor contribution is derived in the Appendix \ref{sec:tensor}. We find that the tensor contribution to the non-linear evolution of density contrast comes through $Z_{ij}$ which is given as
\bea
   & & Z_{ij}
       \equiv N_{ij}
       - 2 \Delta^{-1} \nabla_{(i} N_{j),k}^k
       + {1 \over 2} \Delta^{-2} \left( \nabla_i \nabla_j + \delta_{ij} \Delta \right)
       N^{k\ell}_{\;\;\;,k\ell},
   \nonumber \\
   & & a^2 N_{ij}
       \equiv - {1 \over c^2} \left\{
       u_{,ij} \nabla \cdot {\bf u}
       + {\bf u} \cdot {\bf u}_{,ij}
       - {1 \over 3} \delta_{ij} \left[
       \left( \nabla \cdot {\bf u} \right)^2
       + {\bf u} \cdot \left( \Delta {\bf u} \right) \right]
       \right\},
   \label{Sij}
\eea
where $A_{(ij)} \equiv {1 \over 2} (A_{ij} + A_{ji})$.

With these Newtonian identifications, equations
(\ref{energy-conservation-correction}) and
(\ref{Raychaudhury-eq-correction}) now take the form similar to, respectively, the usual Newtonian energy and momentum conservation equations as,
\bea
   & & \dot \delta
       + {1 \over a} \nabla \cdot {\bf u}
       + {1 \over a} \nabla \cdot \left( \delta {\bf u} \right)
       = {1 \over a} \left( \nabla \delta \right) \cdot
       \left[ 2 \varphi {\bf u}
       - \Delta^{-1} \left( \nabla X + {\bf Y} \right) \right],
   \label{delta-eq} \\
   & & {1 \over a} \nabla \cdot \left( \dot {\bf u}
       + {\dot a \over a} {\bf u} \right)
       + 4 \pi G \varrho \delta
       + {1 \over a^2} \nabla \cdot \left( {\bf u} \cdot \nabla {\bf u} \right)
       = {1 \over a^2} \biggl\{
       - {2 \over 3} \varphi {\bf u} \cdot \nabla
       \left( \nabla \cdot {\bf u} \right)
       + 4 \nabla \cdot \left[
       \varphi \left( {\bf u} \cdot \nabla {\bf u}
       - {1 \over 3} {\bf u} \nabla \cdot {\bf u} \right) \right]
   \nonumber \\
   & & \qquad
       + {2 \over 3} X \nabla \cdot {\bf u}
       + {\bf u} \cdot \left( \nabla X + {\bf Y} \right)
       - \Delta \left[ {\bf u} \cdot \Delta^{-1}
       \left( \nabla X + {\bf Y} \right) \right] \biggr\}
       + 2 {\dot a \over a} {1 \over a} u^{,ij} \Delta^{-1}
       \left( a^2 Z_{ij} \right).
   \label{u-eq}
\eea
These are the general relativistic energy conservation and (divergence of) momentum conservation equations, respectively, of a zero-pressure fluid valid to third order in perturbation. We have assumed a flat background ($\overline K = 0$) and have taken the comoving gauge; to the linear order these equations are known to be also valid on the curved background (that is, in the presence of non-zero spatial curvature $\overline K$; \citep{Bardeen-1980}).

Left-hand-sides are the same as Newtonian perturbation equations; the
zero-pressure Newtonian perturbation equations have quadratic order
non-linearity only, thus these parts are valid to fully nonlinear order in the Newtonian context \citep{Peebles-1980}. Terms in the right-hand sides are third order and purely general relativistic corrections in the zero-pressure fluid. Thus, we have exact relativistic/Newtonian correspondence to the second order perturbations \citep{Noh-Hwang-2004}. We note that our equations are valid to the third order perturbations but in fully relativistic context (e.g., including the super-horizon scale). Compared with our previous
presentation in \citet{Hwang-Noh-2006-MN}, now we have additional ${\bf Y}$ and $Z_{ij}$ terms on the right-hand sides. These correction terms take into account the quadratic-order vector- and tensor-type perturbations simultaneously excited by the linear-order scalar-type perturbations.

We note that while the pure relativistic corrections from the scalar- and vector-type perturbations appear via the linear spatial curvature
perturbation $\varphi ({\bf x},t)$, the correction from tensor-type perturbation involves linear velocity perturbation ${\bf u} ({\bf x},t)$ only, see equation (\ref{Sij}).

\section{Power spectra}

We will present the leading non-linear power spectra of the density and
velocity field. We introduce the velocity gradient variable \citep{Peebles-1980}
\bea
   & & \theta \equiv {1 \over a} \nabla \cdot {\bf u} = - \kappa.
\eea
The solutions of $\delta$ and $\theta$ in the Fourier space are presented in the Appendix \ref{sec:mode-analysis}. The density and velocity power spectra are defined as
\bea
   & & \langle \delta ({\bf k}_1, t) \delta ({\bf k}_2, t) \rangle
       \equiv (2 \pi)^3 \delta^{(3)} ( {\bf k}_1 + {\bf k}_2 )
       P_\delta (k_1, t), \quad
       \langle \theta ({\bf k}_1, t) \theta ({\bf k}_2, t) \rangle
       \equiv (2 \pi)^3 \delta^{(3)} ( {\bf k}_1 + {\bf k}_2 )
       P_\theta (k_1, t).
\eea
We may decompose the non-linear power spectra as $P \equiv P_{11} + P_{22} + P_{13}$, where the subindices $ab$ in $P_{ab}(k)$ stand for two perturbation orders out of which the contribution $P_{ab}(k)$ are calculated. Due to the relativistic/Newtonian correspondence up to second order, the relativistic $P_{11}(k)$ and $P_{22}(k)$ terms are the same as the Newtonian ones, and purely relativistic corrections occur
from $P_{13}$ that involves the third-order solutions. We decompose $P_{13}$ to the relativistic/Newtonian and pure Einstein parts as
$P_{13} = P_{13}^{\rm N} + P_{13}^{\rm E}$, and further decompose
$P_{13}^{\rm E}$ to the scalar-, vector- and tensor-type contributions as $P_{13}^{\rm E} = P_{13}^{\rm ES} + P_{13}^{\rm EV} + P_{13}^{\rm ET}$.

For density power spectrum, we find
\bea
   & & P_{11} (k, t) = |\delta_1 (k, t)|^2,
   \nonumber \\
   & & P_{22} (k, t) = \frac{k^3}{2\pi^2} \int_{-1}^1 {\rm d}x
       \int_0^\infty {\rm d}r
       \left[ \frac{7x + 3r - 10 r x^2}{14 (1+r^2-2rx)} \right]^2
       |\delta_1 (kr, t)|^2
       |\delta_1 (k\sqrt{1+r^2-2rx}, t)|^2,
   \nonumber \\
   & & P_{13}^{\rm N} (k, t) = \frac{k^3}{2\pi^2}
       |\delta_1 (k, t)|^2 \int_0^\infty {\rm d}r
       |\delta_1 (kr, t)|^2 \frac{1}{504 r^3}
       \left[ 2r(6-79 r^2 + 50 r^4 - 21 r^6)
       + 3 (r^2-1)^3(7r^2+2)\ln\left(\frac{|r-1|}{r+1}\right) \right],
   \nonumber \\
   & & P_{13}^{\rm E} (k, t)
       = -\frac{k^3}{2\pi^2} |\delta_1 (k, t)|^2 \frac{k_H(t)}{k}^2
       \int_0^\infty {\rm d}r |\delta_1 (kr, t)|^2
   \nonumber \\
   & & \qquad \times
       \frac{1}{336 r^3}
       \left[ 2r(156 + 631r^2 + 51r^4)
       + 3 (52 + 193 r^2 - 262 r^4 + 17 r^6)
       \ln\left(\frac{|r-1|}{r+1}\right) \right],
   \nonumber \\
   & & P_{13}^{\rm ES} (k, t)
       = \frac{k^3}{2\pi^2} |\delta_1 (k, t)|^2 \frac{k_H(t)}{k}^2
       \int_0^\infty {\rm d}r |\delta_1 (kr, t)|^2
   \nonumber \\
   & & \qquad \times
       \frac{5}{112 r^3}
       \left[ 2 r(-36 - 65 r^2 + 43 r^4)
       - (36 + 53r^2 -46r^4 -43 r^6)
       \ln\left(\frac{|r-1|}{r+1}\right) \right],
   \nonumber \\
   & & P_{13}^{\rm EV} (k, t)
       = \frac{k^3}{2\pi^2} |\delta_1 (k, t)|^2 \frac{k_H(t)}{k}^2
       \int_0^\infty {\rm d}r |\delta_1 (kr, t)|^2
       \frac{5(1-r^2)}{14 r^3}
       \left[ 6(r+2r^3) +(3+5r^2+6r^4)
       \ln\left(\frac{|r-1|}{r+1}\right) \right],
   \nonumber \\
   & & P_{13}^{\rm ET} (k, t)
       = \frac{k^3}{2\pi^2} |\delta_1 (k, t)|^2 \frac{k_H(t)}{k}^2
       \int_0^\infty {\rm d}r |\delta_1 (kr, t)|^2
       \frac{1}{42 r^3}
       \left[ 2r(3 - 2r^2 + 3 r^4) + 3 (r^2 - 1)^2(r^2+1)
       \ln\left(\frac{|r-1|}{r+1}\right) \right].
\eea
For velocity power spectrum we find
\bea
   & & P_{11} (k, t) = |\theta_1 (k, t)|^2,
   \nonumber \\
   & & P_{22} (k, t) = \frac{k^3}{2\pi^2} \int_{-1}^1 {\rm d}x
       \int_0^\infty {\rm d}r
       \left[ \frac{r - 7x + 6 r x^2}{14 (1+r^2-2rx)} \right]^2
       |\delta_1 (kr, t)|^2
       |\delta_1 (k\sqrt{1+r^2-2rx}, t)|^2,
   \nonumber \\
   & & P_{13}^{\rm N} (k, t)
       = \frac{k^3}{2\pi^2} |\delta_1 (k, t)|^2 \frac{k_H(t)}{k}^2
       \int_0^\infty {\rm d}r |\delta_1 (kr, t)|^2
   \nonumber \\
   & & \qquad \times
       \frac{1}{168 r^3}
       \left[2r(6 - 41 r^2 + 2 r^4 - 3 r^6)
       + 3(r^2-1)^3(r^2+2)\ln\left(\frac{|r-1|}{r+1}\right) \right],
   \nonumber \\
   & & P_{13}^{\rm E} (k, t)
       = -\frac{k^3}{2\pi^2} |\delta_1 (k, t)|^2 \frac{k_H(t)}{k}^2
       \int_0^\infty {\rm d}r |\delta_1 (kr, t)|^2
   \nonumber \\
   & & \qquad \times
       \frac{1}{336 r^3}
       \left[ 2r(312 + 527 r^2 + 207 r^4)
       + 3 (104 + 141 r^2 - 314 r^4 + 69 r^6)
       \ln\left(\frac{|r-1|}{r+1}\right) \right],
   \nonumber \\
   & & P_{13}^{\rm ES} (k, t)
       = -\frac{k^3}{2\pi^2} |\delta_1 (k, t)|^2 \frac{k_H(t)}{k}^2
       \int_0^\infty {\rm d}r |\delta_1 (kr, t)|^2
   \nonumber \\
   & & \qquad \times
       \frac{5}{112 r^3}
       \left[ 2 r(72 + 25 r^2 -23 r^4)
       + (72 + r^2 -50r^4 -23 r^6)
       \ln\left(\frac{|r-1|}{r+1}\right) \right],
   \nonumber \\
   & & P_{13}^{\rm EV} (k, t)
       = \frac{k^3}{2\pi^2} |\delta_1 (k, t)|^2 \frac{k_H(t)}{k}^2
       \int_0^\infty {\rm d}r |\delta_1 (kr, t)|^2
       \frac{5(1-r^2)}{14r^3}
       \left[ 2r(6+5r^2) + (6+3r^2+5r^4)
       \ln\left(\frac{|r-1|}{r+1}\right) \right],
   \nonumber \\
   & & P_{13}^{\rm ET} (k, t)
       = \frac{k^3}{2\pi^2} |\delta_1 (k, t)|^2 \frac{k_H(t)}{k}^2
       \int_0^\infty {\rm d}r |\delta_1 (kr, t)|^2
       \frac{1}{21r^3}
       \left[ 2r(3 - 2r^2 + 3 r^4) + 3 (r^2-1)^2(1+r^2 )
       \ln\left(\frac{|r-1|}{r+1}\right) \right].
\eea

\begin{figure}
\centering
\includegraphics[width=0.7\textwidth]{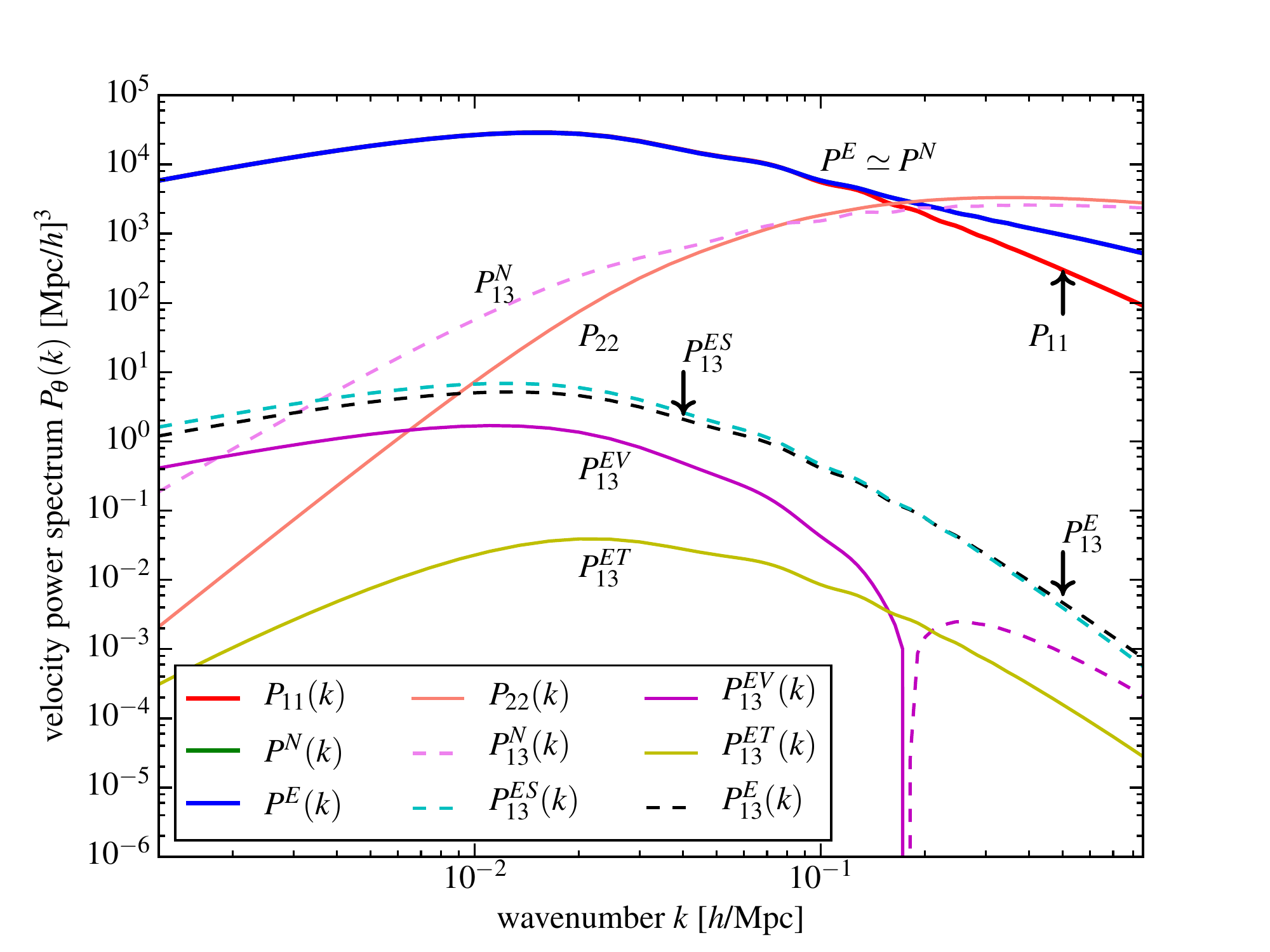}
\caption{
The matter density power spectrum with complete relativistic calculation including all scalar, vector, tensor modes up to third order in perturbation. We assume no linear-order vector and tensor perturbations, but include the vector and tensor perturbations excited due to non-linear mode coupling and their back-reaction on to the scalar modes.
We show various components in the matter power spectrum with different colours, and dashed lines indicate that the components are negative.
The total power spectrum $P^E(k)$ is shown as a solid blue line,
which lies right on top of the Newtonian non-linear matter power spectrum at the same order ($P^{N}(k)$, green line). It is because the total relativistic corrections ($P_{13}^E(k)$, black dashed line) are much smaller than the Newtonian terms ($P_{22}(k)$, orange line,
$P_{13}^N(k)$, pink-dashed line). The relativistic corrections on all scales are dominated by the scalar contribution ($P_{13}^{ES}(k)$ cyan line) over the vector ($P_{13}^{EV}(k)$ magenta line) and tensor ($P_{13}^{ET}(k)$, yellow line) contributions.
}
\label{PS-density}
\end{figure}

\begin{figure}
\centering
\includegraphics[width=0.7\textwidth]{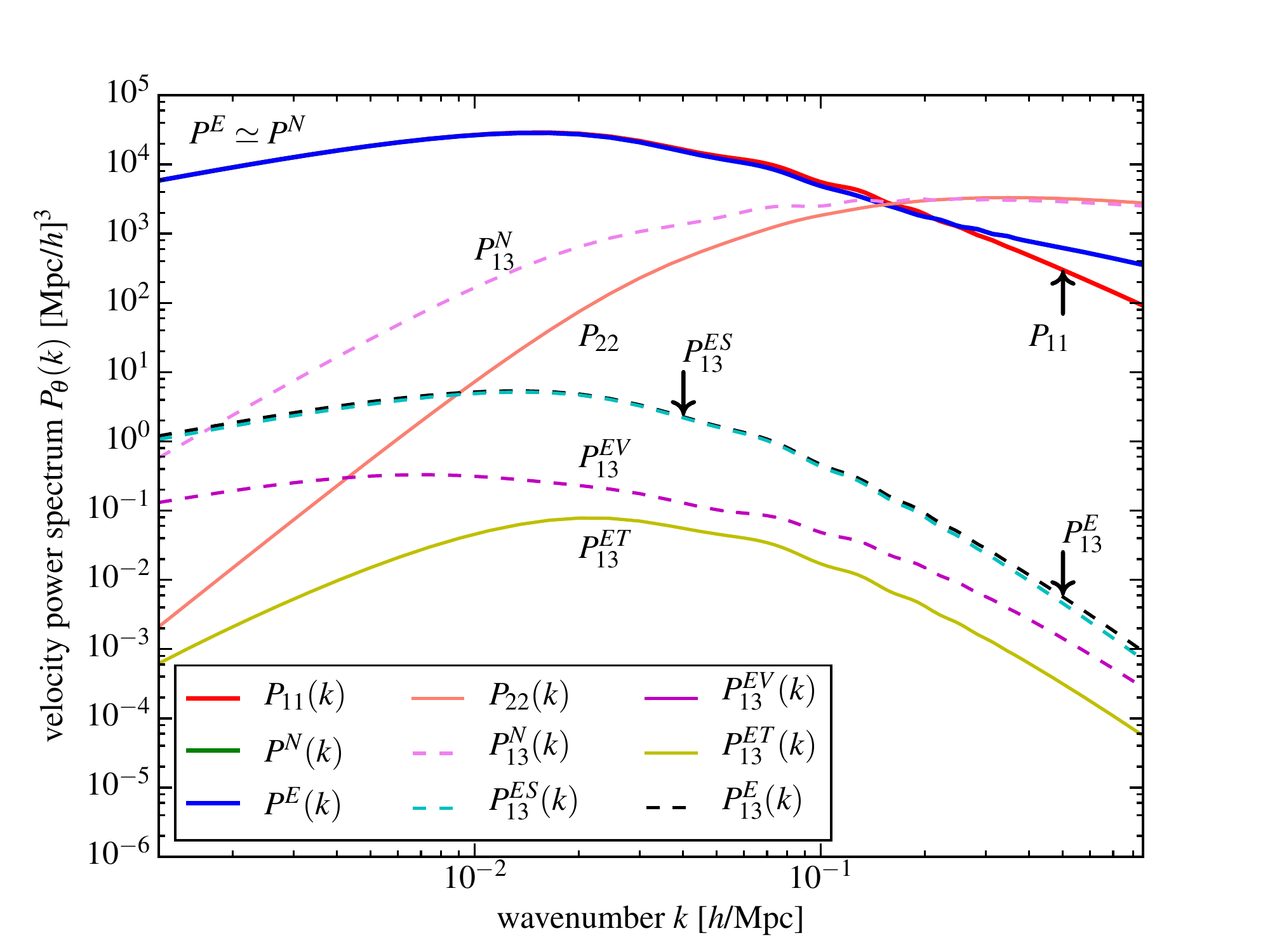}
\caption{
Same as Fig. \ref{PS-density}, but for the third-order power spectrum
of velocity gradient field $\theta\equiv (1/a)\nabla\cdot{\bf u}$. Again, the relativistic corrections are small for all scales, and is dominated by the scalar contribution ($P_{13}^{ES}(k)$).
}
\label{PS-velocity}
\end{figure}

In Figs \ \ref{PS-density} and \ref{PS-velocity}, we present the density and velocity power spectra, respectively, with each contribution represented separately. For the cosmological parameters, we use the maximum likelihood parameters in `WMAP5+BAO+SN' column of table 1 of \citet{wmap5}, and calculate the matter linear power spectrum from the code in \cite{camb}. Both density and velocity power spectra show that, even with the new contributions from induced vector perturbations and tensor perturbations, the pure general relativistic corrections are much smaller than the Newtonian non-linear contributions on all scales.

\section{Discussion}

In this work, we presented the third-order perturbation equations (Section 3) with solutions in Fourier space (Appendix C) and the leading order non-linear density and velocity power spectra (Section 4) of a zero-pressure fluid in the Friedmann background world model. The main new contributions in this work are the following. We include the vector- and tensor-type perturbations accompanied by the linear-order scalar-type perturbation. As these are simultaneously excited by the scalar-type perturbation, it is necessary to include these for a completeness. All these are pure relativistic effects. The full relativistic energy and momentum conservation equations valid to third order perturbation are presented in equations (\ref{delta-eq}) and (\ref{u-eq}). We also present the non-linear power spectra in Figs \ref{PS-density} and \ref{PS-velocity}. The leading nonlinear contributions to density and velocity power spectra via vector- and tensor-type perturbations are smaller than but comparable to the one from pure relativistic scalar-type one. We show these pure relativistic corrections are negligible in all scales. Therefore, as a conclusion we proved that overall effect from the pure Einstein's gravity is negligible in the evolution of zero-pressure fluid during the matter dominated era. That is, even though the system of equations in general relativity is highly non-linear, there is no significant dynamical contributions arising from the non-linearities in general relativity.

As galaxy surveys in the near future such as DESI\footnote{http://desi.lbl.gov}, PFS\footnote{http://sumire.ipmu.jp} will be able to measure primordial non-Gaussianities of order $f_{\rm NL}\simeq 1$, it is important to understand the general relativistic effects (that is expected to be important on large scales; therefore, degenerate with the primordial non-Gaussianity signal) to better accuracy. While we have shown that dynamically induced non-linear corrections in the matter power spectrum are small, there are other general relativistic effects that might be detectable from the large-scale structure observation. For example, \citet{Bartolo-etal-2005} studied effects of primordially generated (i.e., inflationary) non-Gaussianity on the later evolution of the gravitational potential, and \citet{Bruni-etal-2014} likewise
studied the general relativistic effects through the initial conditions encoded during the inflationary era, and similarly considered effects on the non-Gaussianity. Compared with these works emphasizing the second order general relativistic effects imprinted from the early universe, our work concerns pure general relativistic third order non-linearity generated (dynamically induced) from all three (scalar, vector, tensor) types of perturbations during the matter dominated era. In our case, all cosmological parameters are fixed by the $\Lambda$ cold dark matter concordance cosmology, and in that case we have shown that the pure general relativistic effects on the power spectra turn out to be suppressed in all scales. Another general relativistic effect on the observed galaxy power spectrum is the projection effect through the deflection of light coming from galaxies. This effect has been studies extensively to both linear order \citep{Yoo-etal-2009,Yoo-2010,Bonvin-Durrer-2011,Challinor-Lewis-2011,
Jeong-etal-2012,Yoo-2014,Jeong-Schmidt-2015} and to second order 
\citep{Bertacca-etal-2014a,Bertacca-etal-2014b,Yoo-Zaldarriaga-2014}.

There are literatures concerning induced second-order tensor perturbation from the quadratic combinations of linear scalar perturbation (Mollerach et al. 2004; Baumann et al. 2007; Ananda et al. 2007; Sarkar et al. 2008; Arroja et al. 2009; Assadullahi \& Wands 2009; Assadullahi \& Wands 2010; Jedamzik et al. 2010). In this work, we have considered the effects of this induced second order tensor perturbation on the third-order scalar perturbation which is demanded to get the leading order non-linear power spectra. We have included the effect of similarly induced vector perturbation as well. We note an important difference that we considered the comoving gauge in our calculation, whereas {\it all} the previous works on the induced tensor perturbation took the zero-shear (often known as longitudinal, Newtonian or Poisson) gauge.

In this work we only considered the scalar-type perturbation in the linear-order, and assumes no linear order vector- and tensor-type perturbations. The primordially generated gravitational waves (tensor-type perturbation from inflation, for example) may affect the density and velocity power spectra to the non-linear order differently than what has been presented in this work. Although the pure Einstein's gravity effect starts appearing from the third order, the presence of the linear tensor-type perturbation can affect the energy and momentum conservation equation from the second order; see equations 39 and 40 in \citet{Hwang-Noh-2005-third}, and equations 125-128 in \citet{Hwang-Noh-2007-Newtonian} in the presence of multiple component of zero-pressure fluids; although these equations are valid even in the presence of vector-type (rotational) perturbations, as the rotational perturbation only has the decaying mode (in expanding phase) to the linear order, we do not have plausible mechanism to sustain the effect of the rotational perturbations. The effect of primordial tensor-type perturbations on the nonlinear density and velocity power spectra is an open question left for future investigations.

%
%
\section*{Acknowledegments}

JH\ was supported by Basic Science Research Program through the National Research Foundation (NRF) of Korea funded by the Ministry of Science, ICT and Future Planning (No. 2013R1A2A2A01068519).
DJ\ was supported by NSF grant AST-1517363.
HN\ was supported by National Research Foundation of Korea funded by the Korean Government (No.\ 2015R1A2A2A01002791).

%
%
\bibliographystyle{mn2e}


\appendix
%
%
\section{Fully non-linear perturbation equations: Review}
                                          \label{sec:FNL-eqs}

For completeness we present the complete set of fully non-linear scalar- and vector-type perturbation equations in the presence of the additional linear tensor-type perturbation. Our metric convention is (Bardeen 1988, Hwang \& Noh 2013)
\bea
   & & ds^2 = - a^2 \left( 1 + 2 \alpha \right) (d x^0)^2
       - 2 a \chi_{i} d x^0 d x^i
       + a^2 \left[ \left( 1 + 2 \varphi \right) \gamma_{ij}
       + 2 h_{ij} \right] d x^i d x^j,
   \label{metric-PT}
\eea
where spatial indices of $\chi_i$ and $h_{ij}$ are raised and lowered by $\gamma_{ij}$ as the metric. The metric $\gamma_{ij}$ is the comoving part of the three-space metric of the Robertson-Walker spacetime; $6 \overline K$ is the spatial scalar curvature of the background metric $\gamma_{ij}$ (Noh 2014). Here we {\it assume} $a$ to be a function of time only, and $\alpha$, $\varphi$ and $\chi_i$ are functions of space and time with arbitrary amplitude. We include the transverse-tracefree ($h^j_{i|j} \equiv 0 \equiv h^j_j$) tensor-type perturbation {\it only} to the linear order; a vertical bar indicates a covariant derivative based on $\gamma_{ij}$ as the metric. The spatial part of the metric is simple because we already have taken the spatial gauge condition (to fully non-linear order) without losing any generality (Bardeen 1988, Hwang \& Noh 2013).
The energy-momentum tensor of a fluid in the energy-frame is (Ellis 1971, Ellis 1973)
\bea
   & & \widetilde T_{ab} = \widetilde \mu \widetilde u_a \widetilde u_b
       + \widetilde p \left( \widetilde g_{ab}
       + \widetilde u_a \widetilde u_b \right),
   \label{Tab}
\eea
where $\widetilde u_a$ is the normalized fluid four-vector with $\widetilde u^a \widetilde u_a \equiv -1$; $\widetilde \mu$ and $\widetilde p$ are the covariant energy density and pressure, respectively, with tildes indicating covariant quantities. We ignore the anisotropic stress in this work. We introduce (Hwang \& Noh 2013)
\bea
   & & \widetilde u_i
       \equiv a \widehat \gamma {\widehat v_i \over c}, \quad
       \widehat \gamma
       \equiv {1 \over \sqrt{ 1 - {\widehat v^k \widehat v_k
       \over c^2 (1 + 2 \varphi)}}},
   \label{v-PT}
\eea where $\widehat v_i$ is an arbitrary function of space and time with the spatial index raised and lowered by $\gamma_{ij}$ as the metric; $\widehat \gamma$ is the Lorentz factor. The complete set of fully non-linear and exact cosmological perturbation equations without taking the temporal gauge condition is the following; for derivation, see Hwang \& Noh (2013) and Noh (2014).

\noindent
Definition of $\kappa$ (perturbed trace of extrinsic curvature): \bea
   & & \kappa
       \equiv
       3 {\dot a \over a} \left( 1 - {1 \over {\cal N}} \right)
       - {1 \over {\cal N} (1 + 2 \varphi)}
       \left[ 3 \dot \varphi
       + {c \over a^2} \left( \chi^k_{\;\;|k}
       + {\chi^{k} \varphi_{,k} \over 1 + 2 \varphi} \right)
       \right].
   \label{eq1}
\eea
ADM energy constraint:
\bea
   & & - {3 \over 2} \left( {\dot a^2 \over a^2}
       - {8 \pi G \over 3 c^2} \widetilde \mu
       + {\overline K c^2 \over a^2 (1 + 2 \varphi)}
       - {\Lambda c^2 \over 3} \right)
       + {\dot a \over a} \kappa
       + {c^2 \Delta \varphi \over a^2 (1 + 2 \varphi)^2}
   \nonumber \\
   & & \qquad
       = {1 \over 6} \kappa^2
       - {4 \pi G \over c^2} \left( \widetilde \mu + \widetilde p \right)
       \left( \widehat \gamma^2 - 1 \right)
       + {3 \over 2} {c^2 \varphi^{|i} \varphi_{,i} \over a^2 (1 + 2 \varphi)^3}
       - {c^2 \over 4} \overline{K}^i_j \overline{K}^j_i.
   \label{eq2}
\eea
ADM momentum constraint:
\bea
   & & {2 \over 3} \kappa_{,i}
       + {c \over a^2 {\cal N} ( 1 + 2 \varphi )}
       \left[ {1 \over 2} \left( \Delta \chi_i
       + \chi^k_{\;\;|ik} \right)
       - {1 \over 3} \chi^k_{\;\;|ki} \right)
       + {8 \pi G \over c^4} \left( \widetilde \mu + \widetilde p \right)
       a \widehat \gamma^2 \widehat v_i
   \nonumber \\
   & & \qquad
       =
       {c \over a^2 {\cal N} ( 1 + 2 \varphi)}
       \Bigg\{
       \left( {{\cal N}_{,j} \over {\cal N}}
       - {\varphi_{,j} \over 1 + 2 \varphi} \right)
       \left[ {1 \over 2} \left( \chi^{j}_{\;\;|i} + \chi_i^{\;|j} \right)
       - {1 \over 3} \delta^j_i \chi^k_{\;\;|k} \right]
   \nonumber \\
   & & \qquad
       - {\varphi^{,j} \over (1 + 2 \varphi)^2}
       \left( \chi_{i} \varphi_{,j}
       + {1 \over 3} \chi_{j} \varphi_{,i} \right)
       + {{\cal N} \over 1 + 2 \varphi} \nabla_j
       \left[ {1 \over {\cal N}} \left(
       \chi^{j} \varphi_{,i}
       + \chi_{i} \varphi^{|j}
       - {2 \over 3} \delta^j_i \chi^{k} \varphi_{,k} \right) \right]
       \Bigg\}.
   \label{eq3}
\eea
Trace of ADM propagation:
\bea
   & & - 3 \left[ {1 \over {\cal N}}
       \left( {\dot a \over a} \right)^{\displaystyle\cdot}
       + {\dot a^2 \over a^2}
        + {4 \pi G \over 3 c^2} \left( \widetilde \mu + 3 \widetilde p \right)
       - {\Lambda c^2 \over 3} \right]
       + {1 \over {\cal N}} \dot \kappa
       + 2 {\dot a \over a} \kappa
       + {c^2 \Delta {\cal N} \over a^2 {\cal N} (1 + 2 \varphi)}
   \nonumber \\
   & & \qquad
       = {1 \over 3} \kappa^2
       + {8 \pi G \over c^2} \left( \widetilde \mu + \widetilde p \right)
       \left( \widehat \gamma^2 - 1 \right)
       - {c \over a^2 {\cal N} (1 + 2 \varphi)} \left(
       \chi^{i} \kappa_{,i}
       + c {\varphi^{|i} {\cal N}_{,i} \over 1 + 2 \varphi} \right)
       + c^2 \overline{K}^i_j \overline{K}^j_i.
   \label{eq4}
\eea
Tracefree ADM propagation (with linear tensor perturbation):
\bea
   & & \ddot h_{ij}
       + 3 H \dot h_{ij}
       - c^2 {\Delta - 2 \overline K \over a^2} h_{ij}
       +
       \left( {1 \over {\cal N}} {\partial \over \partial t}
       + 3 {\dot a \over a}
       - \kappa
       + {c \chi^{k} \over a^2 {\cal N} (1 + 2 \varphi)} \nabla_k \right)
       \Bigg\{ {c \over a^2 {\cal N} (1 + 2 \varphi)}
   \nonumber \\
   & & \qquad
       \times
       \left[
       {1 \over 2} \left( \chi^i_{\;\;|j} + \chi_j^{\;|i} \right)
       - {1 \over 3} \delta^i_j \chi^k_{\;\;|k}
       - {1 \over 1 + 2 \varphi} \left( \chi^{i} \varphi_{,j}
       + \chi_{j} \varphi^{|i}
       - {2 \over 3} \delta^i_j \chi^{k} \varphi_{,k} \right)
       \right] \Bigg\}
   \nonumber \\
   & & \qquad
       - {c^2 \over a^2 ( 1 + 2 \varphi)}
       \left[ {1 \over 1 + 2 \varphi}
       \left( \nabla^i \nabla_j - {1 \over 3} \delta^i_j \Delta \right) \varphi
       + {1 \over {\cal N}}
       \left( \nabla^i \nabla_j - {1 \over 3} \delta^i_j \Delta \right) {\cal N} \right]
   \nonumber \\
   & & \qquad
       =
       {8 \pi G \over c^2} \left( \widetilde \mu + \widetilde p \right)
       \left[ {\widehat \gamma^2 \widehat v^i \widehat v_j \over c^2 (1 + 2 \varphi)}
       - {1 \over 3} \delta^i_j \left( \widehat \gamma^2 - 1 \right)
       \right]
       + {c^2 \over a^4 {\cal N}^2 (1 + 2 \varphi)^2}
       \Bigg[
       {1 \over 2} \left( \chi^{i|k} \chi_{j|k}
       - \chi_{k|j} \chi^{k|i} \right)
   \nonumber \\
   & & \qquad
       + {1 \over 1 + 2 \varphi} \left(
       \chi^{k|i} \chi_k \varphi_{,j}
       - \chi^{i|k} \chi_j \varphi_{,k}
       + \chi_{k|j} \chi^k \varphi^{|i}
       - \chi_{j|k} \chi^i \varphi^{|k} \right)
       + {2 \over (1 + 2 \varphi)^2} \left(
       \chi^{i} \chi_{j} \varphi^{|k} \varphi_{,k}
       - \chi^{k} \chi_{k} \varphi^{|i} \varphi_{,j} \right) \Bigg]
   \nonumber \\
   & & \qquad
       - {c^2 \over a^2 (1 + 2 \varphi)^2}
       \Bigg[ {3 \over 1 + 2 \varphi}
       \left( \varphi^{|i} \varphi_{,j}
       - {1 \over 3} \delta^i_j \varphi^{|k} \varphi_{,k} \right)
       + {1 \over {\cal N}} \left(
       \varphi^{|i} {\cal N}_{,j}
       + \varphi_{,j} {\cal N}^{|i}
       - {2 \over 3} \delta^i_j \varphi^{|k} {\cal N}_{,k} \right) \Bigg].
   \label{eq5}
\eea
Covariant energy conservation:
\bea
   & &
       \left[ {\partial \over \partial t}
       + {1 \over a ( 1 + 2 \varphi )} \left( {\cal N} \widehat v^k
       + {c \over a} \chi^k \right) \nabla_k \right] \widetilde \mu
       + \left( \widetilde \mu + \widetilde p \right)
       \Bigg\{
       {\cal N} \left( 3 {\dot a \over a} - \kappa \right)
   \nonumber \\
   & & \qquad
       +
       {({\cal N} \widehat v^k)_{|k} \over a (1 + 2 \varphi)}
       + {{\cal N} \widehat v^k \varphi_{,k} \over a (1 + 2 \varphi)^2}
       + {1 \over \widehat \gamma}
       \left[ {\partial \over \partial t}
       + {1 \over a ( 1 + 2 \varphi )} \left( {\cal N} \widehat v^k
       + {c \over a} \chi^k \right) \nabla_k \right] \widehat \gamma \Bigg\}
       = 0.
   \label{eq6}
\eea
Covariant momentum conservation:
\bea
   & & {1 \over a \widehat \gamma}
       \left[ {\partial \over \partial t}
       + {1 \over a ( 1 + 2 \varphi )} \left( {\cal N} \widehat v^k
       + {c \over a} \chi^k \right) \nabla_k \right]
       \left( a \widehat \gamma \widehat v_i \right)
       + \widehat v^k \nabla_i \left( {c \chi_k \over a^2 ( 1 + 2
       \varphi)} \right)
       + {c^2 \over a} {\cal N}_{,i}
       - \left( 1 - {1 \over \widehat \gamma^2} \right) {c^2 {\cal N}
       \varphi_{,i} \over a (1 + 2 \varphi)}
   \nonumber \\
   & & \qquad
       + {1 \over \widetilde \mu + \widetilde p}
       \left\{
       {{\cal N} c^2 \over a \widehat \gamma^2} \widetilde p_{,i}
       + \widehat v_i
       \left[ {\partial \over \partial t}
       + {1 \over a ( 1 + 2 \varphi )} \left( {\cal N} \widehat v^k
       + {c \over a} \chi^k \right) \nabla_k \right] \widetilde p \right\}
       = 0.
   \label{eq7}
\eea
We have
\bea
   & & {\cal N} \equiv \sqrt{
       1 + 2 \alpha + {\chi^k \chi_k \over a^2 (1 + 2 \varphi)}}
       \equiv 1 + \delta {\cal N}, \quad
       \overline{K}^i_j \overline{K}^j_i
       = {1 \over a^4 {\cal N}^2 (1 + 2 \varphi)^2}
       \Bigg\{
       {1 \over 2} \chi^{i|j} \left( \chi_{i|j} + \chi_{j|i} \right)
       - {1 \over 3} \chi^i_{\;\;|i} \chi^j_{\;\;|j}
   \nonumber \\
   & &
       - {4 \over 1 + 2 \varphi} \left[
       {1 \over 2} \chi^i \varphi^{|j} \left(
       \chi_{i|j} + \chi_{j|i} \right)
       - {1 \over 3} \chi^i_{\;\;|i} \chi^j \varphi_{,j} \right]
       + {2 \over (1 + 2 \varphi)^2} \left(
       \chi^{i} \chi_{i} \varphi^{|j} \varphi_{,j}
       + {1 \over 3} \chi^i \chi^j \varphi_{,i} \varphi_{,j} \right) \Bigg\}.
   \label{K-bar-eq}
\eea
As the dimensions we consider
\bea
   & & [\widetilde g_{ab}] = [\widetilde u_a]
       = [a] = [\gamma_{ij}]
       = [\alpha] = [\varphi] = [\chi_i] = [\chi^{(v)}_i]
       = [\widehat v_i/c] = [\widehat v^{(v)}_i/c]
       = [\widehat \gamma] = 1, \quad
       [x^a]
       = [ c dt] = [ d \eta] = L,
   \nonumber \\
   & &
       [\chi] = T, \quad
       [\kappa] = T^{-1}, \quad
       [\widehat v/c] = L, \quad
       [\widetilde T_{ab}] = [\widetilde \mu]
       = [\widetilde \varrho c^2]
       = [\widetilde p], \quad
       [G \widetilde \varrho] = T^{-2}, \quad
       [\Lambda]
       = [\overline K]
       = L^{-2}.
\eea
The perturbation variable $\kappa$ is a perturbed part of the trace of extrinsic curvature ($K^i_i \equiv - 3 \dot a/a + \kappa$). ADM indicates the Arnowitt-Deser-Misner ($3+1$) formulation of Einstein's gravity(Arnowitt, Deser \& Misner 1962). Covariant conservation equations are part of the covariant ($1+3$) formulation equations (Ehlers 1993, Hawking 1966, Ellis 1971, Ellis 1973).

%
%
\section{Tensor contribution}
                                          \label{sec:tensor}

Here we derive the tensor contribution via $\dot h_{ij}$-term in equation (\ref{Raychaudhury-eq-correction}). We consider $\overline K = 0$. We need the tensor-type perturbation generated to the second order with the quadratic combination of linear scalar-type perturbation as the source. The tensor perturbation equation generated by the scalar and vector perturbation to the fully non-linear order is presented in equation 95 of Hwang \& Noh (2013); also see our equation (\ref{eq5}); to the second order, see equation 210 in Noh \& Hwang (2004). To the second order in the comoving gauge considering only the scalar perturbation, from equation (\ref{eq5}), we have
\bea
   & & \ddot h_{ij} + 3 {\dot a \over a} \dot h_{ij} - c^2 {\Delta \over a^2} h_{ij}
       = c^2 s_{ij},
   \label{GW-eq}
\eea
where
\bea
   & & s_{ij}
       = n_{ij}
       - 2 \Delta^{-1} \nabla_{(i} n_{j),k}^k
       + {1 \over 2} \Delta^{-2} \left( \nabla_i \nabla_j + \delta_{ij} \Delta \right)
       n^{k\ell}_{\;\;\;,k\ell},
   \nonumber \\
   & & a^2 n_{ij}
       = \kappa \chi_{,ij}
       - {c^2 \over a^2} \chi^{,k} \chi_{,ijk}
       - 2 \varphi \varphi_{,ij}
       - \varphi_{,i} \varphi_{,j}
       - {1 \over 3} \delta_{ij} \left(
       \kappa \Delta \chi
       - {c^2 \over a^2} \chi^{,k} \Delta \chi_{,k}
       - 2 \varphi \Delta \varphi
       - \varphi^{,k} \varphi_{,k} \right).
   \label{GW-source}
\eea
As we consider $h_{ij}$ sourced by quadratic combinations of linear scalar-type perturbation, $h_{ij}$ can be regarded as the pure second order.

In Einstein-de Sitter model, we have $a \propto t^{2/3} \propto \eta^2$; $\eta$ is the conformal time with $c dt \equiv a d \eta$.
To the linear order, from equations (\ref{eq1}), (\ref{eq3}) and (\ref{eq7}), equations (\ref{eq5}) and (\ref{eq7}), and equation (\ref{eq3}), respectively, we have \bea
   & & \dot \varphi = 0,  \quad
       {1 \over a} \left( a \chi \right)^\cdot = \varphi, \quad
       \kappa = - c^2 {\Delta \over a^2} \chi.
\eea
We can show that the $\varphi$-terms contribute as constants (in time) to $a^2 n_{ij}$ (thus to $a^2 s_{ij}$ as well), and the other terms contribute as $a (\propto \eta^2)$ to $a^2 n_{ij}$; the growing solution of $\chi$ is proportional to $t(\propto a^{3/2})$.
We change equation (\ref{GW-eq}) as
\bea
   & & {\partial^2 h_{ij} \over \partial x^2}
       + {4 \over x} {\partial h_{ij} \over \partial x}
       + h_{ij}
       = - a^2 \Delta^{-1} s_{ij},
\eea
where $x \equiv k \eta$ with the wavenumber $k$ defined as $\Delta = - k^2$.
We set the source part as
\bea
   & & - a^2 \Delta^{-1} s_{ij} \equiv x^2 A_{ij} + B_{ij},
\eea
where $A_{ij}$ and $B_{ij}$ are constants in time.
The homogeneous solution is (Lifshitz 1946)
\bea
   & & h_{ij} \quad
       \propto \quad
       {\eta \over a} j_1 (x), \quad
       {\eta \over a} y_i (x) \quad
       \propto \quad
       {\sin{x} \over x^3} - {\cos{x} \over x^2}, \quad
       {\cos{x} \over x^3} + {\sin{x} \over x^2}.
\eea
Using this, the inhomogeneous solution becomes
\bea
   & & h_{ij} = \left( x^2 - 10 \right) A_{ij} + B_{ij}
       = - a^2 \Delta^{-1} s_{ij} - 10 A_{ij}.
\eea
Thus, we have
\bea
   & & {2 c^2 \over a^2} \chi^{,ij} \dot h_{ij}
       = - {2 c^2 \over a^2} \chi^{,ij}
       \Delta^{-1} \left( a^2 s_{ij} \right)^{\displaystyle\cdot}
       = - 2 H {c^2 \over a^2} \chi^{,ij}
       \Delta^{-1} \left( a^2 s_{ij} - k^2 B_{ij} \right),
\eea
where $s_{ij}$ is in equation (\ref{GW-source}); $B_{ij}$ is terms with $\varphi$ in that equation; we set $a^2 Z_{ij} \equiv a^2 s_{ij} - k^2 B_{ij}$. From these we can show equation (\ref{Sij}).

%
%
\section{Mode-coupling analysis}
                                          \label{sec:mode-analysis}

\subsection{Equations in Fourier space}

In Fourier space with $f ({\bf k}) = \int d^3 x f ({\bf x}) e^{i{\bf k} \cdot {\bf x}}$, equations (\ref{delta-eq}) and (\ref{u-eq}) give
\bea
   & & \dot \delta ({\bf k}, t) + \theta ({\bf k}, t)
       = - {1 \over (2 \pi)^3} \int d^3 q_1 \int d^3 q_2
       \delta^{(3)} ( {\bf k} - {\bf q}_{12} )
       {{\bf q}_{12} \cdot {\bf q}_2 \over q_2^2}
       \delta ({\bf q}_1, t) \theta ({\bf q}_2, t)
   \nonumber \\
   & & \qquad
       + {1 \over (2 \pi)^6} \int d^3 q_1 \int d^3 q_2 \int d^3 q_3
       \delta^{(3)} ( {\bf k} - {\bf q}_{123} )
       {2 {\bf q}_1 \cdot {\bf q}_2 \over q_2^2}
       \delta ({\bf q}_1, t)
       \theta ({\bf q}_2, t) \varphi ({\bf q}_3, t)
   \nonumber \\
   & & \qquad
       - {1 \over (2 \pi)^3} \int d^3 q_1 \int d^3 q_2
       \delta^{(3)} ( {\bf k} - {\bf q}_{12} )
       {1 \over a} \delta ({\bf q}_1, t)
       {1 \over q_2^2} \left[ {\bf q}_1 \cdot {\bf q}_2
       X ({\bf q}_2, t)
       +
       {\bf q}_1 \cdot i {\bf Y} ({\bf q}_2, t)
       \right],
   \\
   & & \dot \theta ({\bf k}, t) + 2 H \theta ({\bf k}, t)
       + 4 \pi G \varrho \delta ({\bf k}, t)
       = - {1 \over (2 \pi)^3} \int d^3 q_1 \int d^3 q_2
       \delta^{(3)} ( {\bf k} - {\bf q}_{12} )
       {{\bf q}_1 \cdot {\bf q}_2 {\bf q}_{12} \cdot {\bf q}_2
       \over q_1^2 q_2^2}
       \theta ({\bf q}_1, t) \theta ({\bf q}_2, t)
   \nonumber \\
   & & \qquad
       + {1 \over (2 \pi)^6} \int d^3 q_1 \int d^3 q_2 \int d^3 q_3
       \delta^{(3)} ( {\bf k} - {\bf q}_{123} )
       {1 \over q_2^2} \left(
       - {2 \over 3} {\bf q}_1 \cdot {\bf q}_2
       + 4 {{\bf q}_{123} \cdot {\bf q}_1 {\bf q}_1 \cdot {\bf q}_2
       \over q_1^2}
       - {4 \over 3} {\bf q}_{123} \cdot {\bf q}_2
       \right)
       \theta ({\bf q}_1, t) \theta ({\bf q}_2, t)
       \varphi ({\bf q}_3, t)
   \nonumber \\
   & & \qquad
       + {1 \over (2 \pi)^3} \int d^3 q_1 \int d^3 q_2
       \delta^{(3)} ( {\bf k} - {\bf q}_{12} )
       {1 \over a} \theta ({\bf q}_1, t)
       \left\{
       {2 \over 3} X ({\bf q}_2, t)
       + {1 \over q_1^2} \left( 1 - {q_{12}^2 \over q_2^2} \right)
       \left[ {\bf q}_1 \cdot {\bf q}_2 X ({\bf q}_2, t)
       + {\bf q}_1 \cdot i {\bf Y} ({\bf q}_2, t) \right]
       \right\}
   \nonumber \\
   & & \qquad
       - {1 \over (2 \pi)^3} \int d^3 q_1 \int d^3 q_2
       \delta^{(3)} ( {\bf k} - {\bf q}_{12} )
       {q_1^i q_1^j \over q_1^2 q_2^2}
       2 H \theta ({\bf q}_1, t) a^2 Z_{ij} ({\bf q}_2, t),
\eea
where equations (\ref{XY}) and (\ref{Sij}) give
\bea
   & & X ({\bf k}, t)
       = {1 \over (2 \pi)^3} \int d^3 q_1 \int d^3 q_2
       \delta^{(3)} ( {\bf k} - {\bf q}_{12} )
       \left(
       2 - {{\bf q}_1 \cdot {\bf q}_2 \over q_1^2}
       + {3 \over 2} {{\bf q}_{12} \cdot {\bf q}_2 \over q_{12}^2}
       {{\bf q}_1 \cdot {\bf q}_2 \over q_1^2}
       + {3 \over 2} {{\bf q}_{12} \cdot {\bf q}_1 \over q_{12}^2}
       {q_2^2 \over q_1^2}
       \right)
       a \theta ({\bf q}_1, t) \varphi ({\bf q}_2, t),
   \nonumber \\
   & & i {\bf Y} ({\bf k}, t)
       = {1 \over (2 \pi)^3} \int d^3 q_1 \int d^3 q_2
       \delta^{(3)} ( {\bf k} - {\bf q}_{12} )
       \left(
       {\bf q}_2 {{\bf q}_1 \cdot {\bf q}_2 \over q_1^2}
       + {\bf q}_1 {q_2^2 \over q_1^2}
       - {\bf q}_{12} {{\bf q}_{12} \cdot {\bf q}_2 \over q_{12}^2}
       {{\bf q}_1 \cdot {\bf q}_2 \over q_1^2}
       - {\bf q}_{12} {{\bf q}_{12} \cdot {\bf q}_1 \over q_{12}^2}
       {q_2^2 \over q_1^2}
       \right)
   \nonumber \\
   & & \qquad
       \times
       2 a \theta ({\bf q}_1, t) \varphi ({\bf q}_2, t),
   \nonumber \\
   & & Z_{ij} ({\bf k}, t)
       = N_{ij} ({\bf k}, t)
       -2 {k_k k_{(i} \over k^2} N^k_{j)} ({\bf k}, t)
       + {1 \over 2} \left(
       {k_i k_j \over k^2} + \delta_{ij} \right)
       {k^k k^\ell \over k^2} N_{k\ell} ({\bf k}, t),
   \nonumber \\
   & & N_{ij} ({\bf k}, t)
       = {1 \over (2 \pi)^3} \int d^3 q_1 \int d^3 q_2
       \delta^{(3)} ( {\bf k} - {\bf q}_{12} )
       \left( {1 \over 3} \delta_{ij}
       - {q_{2i} q_{2j} \over q_2^2} \right)
       \left( 1 + {{\bf q}_1 \cdot {\bf q}_2 \over q_1^2} \right)
       {1 \over c^2} \theta ({\bf q}_1, t) \theta ({\bf q}_2, t).
\eea
It is convenient to have
\bea
   & & {\bf k} X({\bf k},t) + i{\bf Y}({\bf k},t)
       = {1 \over (2\pi)^3} \int d^3q_1\int d^3q_2
       \delta^{(3)}({\bf k}-{\bf q}_{12})
       a\theta({\bf q}_1,t)\varphi({\bf q}_2,t)
   \nonumber \\
   & & \qquad
       \times
       \left[
       {\bf q}_{12} \left( 2
       -\frac{{\bf q}_1\cdot{\bf q}_2}{q_1^2}
       - \frac12 (4y-3) \frac{({\bf q}_{12} \cdot {\bf q}_2)
       ({\bf q}_1 \cdot {\bf q}_2)
       + ({\bf q}_{12} \cdot {\bf q}_1) q_2^2}{q_{12}^2 q_1^2} \right)
       + 2y \left( {\bf q}_2 \frac{{\bf q}_1 \cdot {\bf q}_2}{q_1^2}
       + {\bf q}_1 \frac{q_2^2}{q_1^2} \right) \right],
   \\
   & &
	   q_1^iq_1^j Z_{ij} ({\bf k}, t)
       =
       {1 \over (2 \pi)^3} \int d^3 q_2 \int d^3 q_3
       \delta^{(3)} ( {\bf k} - {\bf q}_{23} )
		\theta({\bf q}_2,t)\theta({\bf q}_3,t)
		\left(1+\frac{{\bf q}_2\cdot{\bf q}_3}{q_2^2}\right) z
   \nonumber \\
   & & \qquad
       \times
	   \biggl\{
       {1 \over 3} q_1^2
       - {({\bf q}_1\cdot{\bf q}_3)^2 \over q_{3}^2}
       -
		2 {{\bf q}_1\cdot{\bf k} \over k^2}
       \left( {1 \over 3} {\bf q}_1\cdot{\bf k}
       - {{\bf q}_1\cdot{\bf q}_3 {\bf k}\cdot{\bf q}_3 \over q_{3}^2} \right)
       +
       {1 \over 2} \left(
       {({\bf q}_1\cdot{\bf k})^2 \over k^2} + q_1^2 \right)
       \left( {1 \over 3}
       - {({\bf k}\cdot{\bf q}_3)^2 \over k^2q_{3}^2} \right)
	   \biggl\},
\eea
where we introduced $y$ and $z$ indicating the vector- and tensor-type contributions, respectively; by setting $y=1=z$ we are considering these contributions simultaneously excited by the linear scalar-type perturbation.

\subsection{Linear solutions}

We set the linear density and velocity fields as
\bea
   & & \delta_1({\bf k},t) = D(t)\delta_1({\bf k}),
   \nonumber \\
   & &
       \theta_1({\bf k},t) = -H(t)f(t)D(t)\delta_1({\bf k}),
\eea
with $f\equiv d\ln D/d\ln a$ satisfying, from equation (\ref{u-eq})
\bea
   & & \frac{d(HfD)}{dt} + 2 H^2fD - 4 \pi G \varrho D = 0,
\eea
or, in a well known form (Lifshitz 1946, Bardeen 1980, Peebles 1980)
\bea
   & & \ddot D + 2 H \dot D - 4 \pi G \varrho D = 0,
   \label{D-eq}
\eea
where $H \equiv \dot a/a$.
This linear order equation is valid for general $\overline K$ and $\Lambda$ in the background (Bardeen 1980), and is also available in the Newtonian context (Bonnor 1957, Peebles 1980). Our equation is valid in the comoving gauge and (only to the linear order) the same equation is also derived in the synchronous gauge (Lifshitz 1946). Considering general $\overline K$ and $\Lambda$, equation (\ref{D-eq}) can be written as (Hwang 1994)
\bea
   & & {1 \over a \dot a} \left[
       \dot a^2 \left( {D \over H}
       \right)^{\displaystyle\cdot}
       \right]^{\displaystyle\cdot}
       = 0,
\eea
with general solution (Heath 1977, section 10 in Peebles 1980)
\bea
   & & \delta ({\bf k}, t)
       = \left( k^2 - 3 \overline K \right) C ({\bf k}) H \int^t
       {dt \over \dot a^2}.
\eea
The transient (in expanding phase) mode proportional to $H$ is absorbed in the lower bound of integration. The coefficient $C ({\bf k})$ is normalized so that we have (Hwang 1994)
\bea
   & & \varphi ({\bf k}, t)
       = C ({\bf k}) \left( 1 + \overline K H \int^t
       {dt \over \dot a^2} \right).
   \label{varphi-linear-sol}
\eea
As mentioned below equation (\ref{varphi-eq2}) $\varphi$ is constant for vanishing $\overline K$. In Einstein-de Sitter model we have $\delta \propto a, a^{-3/2}$, thus the growing mode gives $D \propto a$. Solutions with general $\Lambda$ can be found in Edwards \& Heath (1976), Heath (1977), section 13 in Peebles (1980), Weinberg (1987), and Heath (1989).

\subsection{Nonlinear solutions}

We expand the higher order density and velocity fields as
\bea
   & & \delta_n ({\bf k}, t) =
       \frac{D^n (t)}{(2\pi)^{3(n-1)}}
       \int d^3 q_1 \cdots \int d^3 q_n
       \delta^{(3)} ({\bf k}-{\bf q}_{12 \cdots n})
       F_n ({\bf q}_1, \cdots, {\bf q}_n, t)
       \delta_1 ({\bf q}_1)
       \cdots
       \delta_1 ({\bf q}_n),
   \nonumber \\
   & & \theta_n ({\bf k}, t) =
       -\frac{H(t) f(t) D^n(t)}{(2\pi)^{3(n-1)}}
       \int d^3 q_1 \cdots \int d^3 q_n
       \delta^{(3)} ({\bf k} - {\bf q}_{12 \cdots n})
       G_n ({\bf q}_1, \cdots, {\bf q}_n, t)
       \delta_1({\bf q}_1) \cdots \delta_1({\bf q}_n),
   \label{kernel-ansatz}
\eea
with $F_n$ and $G_n$, the density and velocity kernels, respectively. We have $F_1 = 1 = G_1$. As the pure relativistic contribution appears from the third order, the linear and second order solutions are the same as the usual Newtonian ones.

Now we consider Einstein-de Sitter model. We have $D \propto a$, thus $f = 1$ and $\theta = - H \delta$. To the linear order equation (\ref{varphi-eq2}) gives
\bea
   & & \varphi_1 ({\bf k})
       = \frac{a^2}{k^2 c^2}
       \left( 4 \pi G \varrho \delta - H \theta \right)
       = \frac52 \frac{a^2 H^2}{k^2 c^2} \delta_1 ({\bf k}, t).
\eea
We then calculate the right hand sides of the energy and the momentum conservation equations, to third order in $\delta({\bf k})$ as
\bea
   & & \dot \delta ({\bf k}, t) + \theta ({\bf k}, t)
       = - {1 \over (2\pi)^3} \int d^3 q_1 \int d^3 q_2
       \delta^{(3)} ({\bf k} - {\bf q}_{12})
       \delta ({\bf q}_1, t) \theta ({\bf q}_2, t)
       {{\bf q}_{12} \cdot {\bf q}_2 \over q_2^2}
   \nonumber \\
   & & \qquad
       + {a^2H^2 \over c^2} {1 \over (2\pi)^6}
       \int d^3 q_1 \int d^3 q_2 \int d^3 q_3
       \delta^{(3)} ({\bf k} - {\bf q}_{123})
       \delta ({\bf q}_1, t)
       \theta ({\bf q}_2, t) \delta ({\bf q}_3, t)
       {5 {\bf q}_1 \cdot {\bf q}_2 \over q_2^2 q_3^2}
   \nonumber \\
   & & \qquad
       - \frac52 {a^2 H^2 \over c^2} {1 \over (2\pi)^6}
       \int d^3 q_1 \int d^3 q_2 \int d^3 q_3
       \delta^{(3)} ({\bf k} - {\bf q}_{123})
		\delta ({\bf q}_1, t)
		\theta ({\bf q}_2, t) \delta ({\bf q}_3, t)
       {1 \over q_{23}^2 q_3^2}
   \nonumber \\
   & & \qquad
       \times
	   \left[ {\bf q}_1 \cdot {\bf q}_{23}
	   \left(
	   2 - \frac{{\bf q}_2 \cdot {\bf q}_3}{q_2^2}
	   - \frac12 (4y-3) {{\bf q}_{23} \cdot {\bf q}_3
       {\bf q}_2 \cdot {\bf q}_3
	   + {\bf q}_{23} \cdot {\bf q}_{2}q_3^2
       \over q_{23}^2 q_2^2} \right)
	   + 2y \left( {\bf q}_1 \cdot {\bf q}_3
       {{\bf q}_2 \cdot {\bf q}_3 \over q_2^2}
       + {\bf q}_1 \cdot {\bf q}_2 {q_3^2 \over q_2^2} \right)
       \right],
   \\
   & & \dot \theta ({\bf k}, t) + 2 H \theta ({\bf k}, t)
       + 4 \pi G \varrho \delta ({\bf k}, t)
       = - {1 \over (2 \pi)^3} \int d^3 q_1 \int d^3 q_2
       \delta^{(3)} ( {\bf k} - {\bf q}_{12} )
       \theta ({\bf q}_1, t) \theta ({\bf q}_2, t)
       {{\bf q}_1 \cdot {\bf q}_2 {\bf q}_{12} \cdot {\bf q}_2
       \over q_1^2 q_2^2 }
   \nonumber \\
   & & \qquad
       + \frac52 {a^2 H^2 \over c^2} {1 \over (2 \pi)^6} \int d^3 q_1 \int d^3 q_2 \int d^3 q_3
       \delta^{(3)} ( {\bf k} - {\bf q}_{123} )
       \theta ({\bf q}_1, t) \theta ({\bf q}_2, t)
       \delta ({\bf q}_3, t)
   \nonumber \\
   & & \qquad
       \times
       {1 \over q_2^2q_3^2} \left(
       - {2 \over 3} {\bf q}_1 \cdot {\bf q}_2
       + 4 {{\bf q}_{123} \cdot {\bf q}_1 {\bf q}_1 \cdot {\bf q}_2
       \over q_1^2}
       - {4 \over 3} {\bf q}_{123} \cdot {\bf q}_2
       \right)
   \nonumber \\
   & & \qquad
       + \frac52 {a^2 H^2 \over c^2} {1 \over (2 \pi)^6}
       \int d^3 q_1 \int d^3 q_2 \int d^3 q_3
       \delta^{(3)} ( {\bf k} - {\bf q}_{123} )
       \theta ({\bf q}_1, t)
	   \theta({\bf q}_2, t)
	   \delta({\bf q}_3, t)
   \nonumber \\
   & & \qquad
       \times
       \frac{1}{q_3^2}
       \Bigg\{
       {2 \over 3} \left( 2 - \frac{{\bf q}_2 \cdot {\bf q}_3}{q_2^2}
       + \frac32 {{\bf q}_{23} \cdot {\bf q}_3
       {\bf q}_2 \cdot {\bf q}_3
       + {\bf q}_{23} \cdot {\bf q}_2 q_3^2
       \over q_{23}^2 q_2^2} \right)
       + {1 \over q_1^2} \left( 1 - {q_{123}^2 \over q_{23}^2} \right)
   \nonumber \\
   & & \qquad
       \times
	   \left[ {\bf q}_1 \cdot {\bf q}_{23}
	   \left(
	   2 - \frac{{\bf q}_2 \cdot {\bf q}_3}{q_2^2}
	   - \frac12 (4y - 3) {{\bf q}_{23} \cdot {\bf q}_3
       {\bf q}_2 \cdot {\bf q}_3
       + {\bf q}_{23} \cdot {\bf q}_2 q_3^2
       \over q_{23}^2 q_2^2} \right)
	   + 2y \left( {\bf q}_1 \cdot {\bf q}_3
       {{\bf q}_2 \cdot {\bf q}_3 \over q_2^2}
       + {\bf q}_1 \cdot {\bf q}_2 {q_3^2 \over q_2^2} \right)
       \right]
       \Bigg\}
   \nonumber \\
   & & \qquad
       - 2 {a^2 H \over c^2} {1 \over (2 \pi)^6} \int d^3 q_1 \int d^3 q_2 \int d^3 q_3
       \delta^{(3)} ( {\bf k} - {\bf q}_{123} )
       \theta ({\bf q}_1, t) \theta ({\bf q}_2, t)
       \theta ({\bf q}_3, t)
       {1 \over q_1^2 q_{23}^2}
	   \left( 1 + \frac{{\bf q}_2 \cdot {\bf q}_3}{q_2^2} \right) z
   \nonumber \\
   & & \qquad
       \times
	   \left[
       {1 \over 3} q_1^2
       - {({\bf q}_1 \cdot {\bf q}_3)^2 \over q_{3}^2}
       - 2 {{\bf q}_1 \cdot {\bf q}_{23} \over q_{23}^2}
       \left( {1 \over 3} {\bf q}_1\cdot{\bf q}_{23}
       - {{\bf q}_1 \cdot {\bf q}_3 {\bf q}_{23} \cdot {\bf q}_3
       \over q_{3}^2} \right)
       + {1 \over 2} \left(
       {({\bf q}_1 \cdot {\bf q}_{23})^2
       \over q_{23}^2} + q_1^2 \right)
       \left( {1 \over 3}
       - {({\bf q}_{23} \cdot {\bf q}_3)^2 \over q_{23}^2 q_{3}^2} \right)
	   \right].
\eea

Now, we use the kernel ansatz in equation (\ref{kernel-ansatz}) to translate the equation above to the equations for the second and third order kernels. Using $k_H \equiv aH/c$, we have
\bea
   & & {1 \over H} {d F_2 \over dt} + 2 F_2 - G_2
       = \frac{{\bf q}_{12} \cdot {\bf q}_2}{q_2^2}
       \equiv A_2,
   \nonumber \\
   & & - {1 \over H} {d G_2 \over dt}
       - G_2 + {4 \pi G \varrho \over H^2}
       \left( F_2 - G_2 \right)
       = - \frac12 \frac{q_{12}^2 {\bf q}_1 \cdot {\bf q}_2}{q_1^2 q_2^2}
       \equiv B_2,
   \label{G2} \\
   & & {1 \over H} {d F_3 \over dt} + 3 F_3 - G_3
       = \frac{{\bf q}_{123} \cdot {\bf q}_3}{q_3^2}
       F_2({\bf q}_1, {\bf q}_2)
       + \frac{{\bf q}_{123} \cdot {\bf q}_{23}}{q_{23}^2}
       G_2({\bf q}_2, {\bf q}_3)
       - k_H^2
       \frac{5 {\bf q}_1 \cdot {\bf q}_2}{q_2^2 q_3^2}
       + \frac52 k_H^2 {1 \over q_{23}^2q_3^2}
   \nonumber \\
   & & \qquad
       \times
	   \left[ {\bf q}_1 \cdot {\bf q}_{23}
	   \left( 2 - \frac{{\bf q}_2 \cdot {\bf q}_3}{q_2^2}
	   - \frac12 (4y-3) {{\bf q}_{23} \cdot {\bf q}_3
       {\bf q}_2 \cdot {\bf q}_3
       + {\bf q}_{23} \cdot {\bf q}_2 q_3^2
       \over q_{23}^2 q_2^2} \right)
	   + 2y \left( {\bf q}_1 \cdot {\bf q}_3
       {{\bf q}_2 \cdot {\bf q}_3 \over q_2^2}
       + {\bf q}_1 \cdot {\bf q}_2 {q_3^2 \over q_2^2} \right)
       \right]
   \nonumber \\
   & & \qquad
       \equiv A_3 + k_H^2 C_3,
   \nonumber \\
   & & - {1 \over H} {d G_3 \over dt}
       - 2 G_3 + {4 \pi G \varrho \over H^2}
       \left( F_3 - G_3 \right)
       = - \frac{q_{123}^2 {\bf q}_1 \cdot {\bf q}_{23}}{q_1^2 q_{23}^2}
       G_2 ({\bf q}_2, {\bf q}_3)
   \nonumber \\
   & & \qquad
       + \frac52 k_H^2
       {1 \over q_2^2 q_3^2} \left(
       - {2 \over 3} {\bf q}_1 \cdot {\bf q}_2
       + 4 {{\bf q}_{123} \cdot {\bf q}_1 {\bf q}_1 \cdot {\bf q}_2
       \over q_1^2}
       - {4 \over 3} {\bf q}_{123} \cdot {\bf q}_2
       \right)
   \nonumber \\
   & & \qquad
       +\frac52 k_H^2
       \frac{1}{q_3^2}
       \Bigg\{
       {2 \over 3} \left( 2 - \frac{{\bf q}_2 \cdot {\bf q}_3}{q_2^2}
       + \frac32 {{\bf q}_{23} \cdot {\bf q}_3
       {\bf q}_2 \cdot {\bf q}_3
       + {\bf q}_{23} \cdot {\bf q}_2 q_3^2
       \over q_{23}^2 q_2^2} \right)
       + {1 \over q_1^2} \left( 1 - {q_{123}^2 \over q_{23}^2} \right)
   \nonumber \\
   & & \qquad
       \times
	   \left[ {\bf q}_1 \cdot {\bf q}_{23}
	   \left(
	   2 - \frac{{\bf q}_2 \cdot {\bf q}_3}{q_2^2}
	   - \frac12 (4y-3) {{\bf q}_{23} \cdot {\bf q}_3
       {\bf q}_2 \cdot {\bf q}_3
       + {\bf q}_{23} \cdot {\bf q}_2 q_3^2
       \over q_{23}^2 q_2^2} \right)
	   + 2y \left( {\bf q}_1 \cdot {\bf q}_3
       {{\bf q}_2 \cdot {\bf q}_3 \over q_2^2}
       + {\bf q}_1 \cdot {\bf q}_2 {q_3^2 \over q_2^2} \right)
       \right]
       \Bigg\}
   \nonumber \\
   & & \qquad
       + 2 k_H^2 \frac{1}{q_1^2 q_{23}^2}
       \left( 1 + \frac{{\bf q}_2 \cdot {\bf q}_3}{q_2^2}\right)
       z
   \nonumber \\
   & & \qquad
       \times
	   \left[ {1 \over 3} q_1^2
       - {({\bf q}_1 \cdot {\bf q}_3)^2 \over q_{3}^2}
       - 2 {{\bf q}_1 \cdot {\bf q}_{23} \over q_{23}^2}
       \left( {1 \over 3} {\bf q}_1 \cdot {\bf q}_{23}
       - {{\bf q}_1 \cdot {\bf q}_3 {\bf q}_{23} \cdot {\bf q}_3
       \over q_{3}^2} \right)
       + {1 \over 2} \left(
       {({\bf q}_1 \cdot {\bf q}_{23})^2
       \over q_{23}^2} + q_1^2 \right)
       \left( {1 \over 3}
       - {({\bf q}_{23}\cdot{\bf q}_3)^2 \over q_{23}^2q_{3}^2} \right)
       \right]
   \nonumber \\
   & & \qquad
       \equiv B_3 + k_H^2 D_3,
   \label{G3}
\eea
where $A_2, B_2, \dots$ are constants in time. Using $k_H^2 \propto a^{-1}$, the second and third order kernels have the solutions
\bea
   & & F_2 = {1 \over 7} \left( 5 A_2 - 2 B_2 \right), \quad
       G_2 = {1 \over 7} \left( 3 A_2 - 4 B_2 \right),
   \nonumber \\
   & & F_3 = {7 \over 18} A_3 - {1 \over 9} B_3
       + {1 \over 7} k_H^2 \left( 5 C_3 - 2 D_3 \right), \quad
       G_3 = {1 \over 6} A_3 - {1 \over 3} B_3
       + {1 \over 7} k_H^2 \left( 3 C_3 - 4 D_3 \right).
\eea
We need symmetrization over wavenumber indices of $F_2 ({\bf q}_1,{\bf q}_2)$, $G_2 ({\bf q}_1,{\bf q}_2)$, $F_3 ({\bf q}_1,{\bf q}_2,{\bf q}_3, t)$ and $G_3 ({\bf q}_1,{\bf q}_2,{\bf q}_3, t)$.
The pure relativistic corrections are contained in the coefficients of $k_H^2$ terms.

\bsp
\label{lastpage}


\begin{thebibliography}{99}

\bibitem[\protect\citeauthoryear{{Ananda},}{{Ananda}}{2007}]{Ananda-etal}
         Ananda, K. N., Clarkson, C., Wands, D., 2007, Phys. Rev. D, 75, 123518

\bibitem[\protect\citeauthoryear{{Arnowitt} et~al.,}{{Arnowitt} et~al.}{1962}]{ADM}
         Arnowitt R., Deser S., Misner C. W., 1962, in Witten L., ed., {Gravitation: an introduction to current research}. Wiley, New York

\bibitem[\protect\citeauthoryear{{Arroja},}{{Arroja}}{2009}]{Arroja-etal}
         Arroja, F., Assadullahi, H., Koyama, K., Wands, D., 2009, Phys. Rev. D, 80, 123526

\bibitem[\protect\citeauthoryear{{Assadullahi},}{{Assadullahi}}{2009}]{Assadullahi-etal-2009}
         Assadullahi, H., Wands, D., 2009, Phys. Rev. D, 79, 083511

\bibitem[\protect\citeauthoryear{{Assadullahi},}{{Assadullahi}}{2010}]{Assadullahi-etal-2010}
         Assadullahi, H., Wands, D., 2010, Phys. Rev. D, 81, 023527

\bibitem[\protect\citeauthoryear{{Bardeen},}{{Bardeen}}{1980}]{Bardeen-1980}
         Bardeen J. M., 1980, Phys. Rev. D, 22, 1882

\bibitem[\protect\citeauthoryear{{Bardeen},}{{Bardeen}}{1988}]{Bardeen-1988}
         Bardeen J. M., 1988, in Fang L., Zee A., eds, {Particle Physics and Cosmology}. Gordon and Breach, London, page 1

\bibitem[\protect\citeauthoryear{{Bartolo} et al.,}{{Bartolo} et al.}{2005}]{Bartolo-etal-2005}
		Bartolo, N., Matarrese, S., Riotto, A., 2005, J. Cosmol. Astropart. Phys., 0510, 010

\bibitem[\protect\citeauthoryear{{Baumann},}{{Baumann}}{2007}]{GW-Steinhardt}
         Baumann, D., Steinhardt, P., Takahashi, K., Ichiki, K., 2007,
                     Phys. Rev. D, 76, 084019

\bibitem[\protect\citeauthoryear{{Bertacca} et al.,}{{Bertacca} et al.}{2014a}]{Bertacca-etal-2014a}
		Bertacca, D., Maartens, R., Clarkson, C., 2014, J. Cosmol. Astropart. Phys., 9, 037

\bibitem[\protect\citeauthoryear{{Bertacca} et al.,}{{Bertacca} et al.}{2014b}]{Bertacca-etal-2014b}
		Bertacca, D., Maartens, R., Clarkson, C., 2014, J. Cosmol. Astropart. Phys., 11, 013


\bibitem[\protect\citeauthoryear{{Bonnor},}{{Bonnor}}{1957}]{Bonnor-1957}
         Bonnor W. B., 1957, MNRAS, {117}, 104

\bibitem[\protect\citeauthoryear{{Bonvin} and {Durrer},}{{Bonvin} \& {Durrer}} {2011}]{Bonvin-Durrer-2011}
			Bonvin C., Durrer R., 2011, Phys. Rev. D, 86, 063505

\bibitem[\protect\citeauthoryear{{Bruni} et al.,}{{Bruni} et al.}{2014}]{Bruni-etal-2014}
		Bruni, M., Hidalgo, J. C., Wands, D., 2014, ApJ, 793, L11

\bibitem[\protect\citeauthoryear{{Challinor} and {Lewis},}{{Challinor} \& {Lewis}} {2011}]{Challinor-Lewis-2011}
			Challinor A., Lewis A., 2011, Phys. Rev. D, 84, 043516

\bibitem[\protect\citeauthoryear{{Edward},}{{Edward}}{1976}]{Edward-Heath-1976}
         Edward D., Heath D. J., 1976, Ap\&SS, {41}, 183

\bibitem[\protect\citeauthoryear{{Ehlers}}{{Ehlers}}{1993}]{Ehlers-1961}
         Ehlers J., 1993, in, Gen. Rel. Gravit., {25}, 1225

\bibitem[\protect\citeauthoryear{{Ellis},}{{Ellis} et~al.}{1971}]{Ellis-1971}
         Ellis G. F. R., 1971, in Sachs R. K., ed., Proc. Int. Summer School Phys. Enrico Fermi Course 47, General Relativity and Cosmology, Academic Press, New York

\bibitem[\protect\citeauthoryear{{Ellis},}{{Ellis}}{1973}]{Ellis-1973}
         Ellis G. F. R., 1973, in Schatzmann E., ed., Cargese Lectures in Physics. Gordon and Breach, New York

\bibitem[\protect\citeauthoryear{{Friedmann},}{{Friedmann}}{1922}]{Friedmann-1922}
         Friedmann A. A., 1922, Z. Phys., 10, 377 

\bibitem[\protect\citeauthoryear{{Hawking},}{{Hawking}}{1966}]{Hawking-1966}
         Hawking S. W., 1966, ApJ, 145, 544

\bibitem[\protect\citeauthoryear{{Heath},}{{Heath}}{1977}]{Heath-1977}
         Heath D. J., 1977, MNRAS, {179}, 351

\bibitem[\protect\citeauthoryear{{Heath},}{{Heath}}{1989}]{Heath-1989}
         Heath D. J., 1989, Ap\&SS, {154}, 207

\bibitem[\protect\citeauthoryear{{Hwang},}{{Hwang}}{1994}]{Hwang-1994}
         Hwang J., 1994, ApJ, {427}, 533

\bibitem[\protect\citeauthoryear{{Hwang} and {Noh},}{{Hwang} \& {Noh}}{2005}]{Hwang-Noh-2005-third}
         Hwang J., Noh H., 2005, Phys. Rev. D, {72}, 044012

\bibitem[\protect\citeauthoryear{{Hwang} and {Noh},}{{Hwang} \& {Noh}}{2006}]{Hwang-Noh-2006-MN}
         Hwang J., Noh H., 2006, MNRAS, 367, 1515

\bibitem[\protect\citeauthoryear{{Hwang} and {Noh},}{{Hwang} \& {Noh}}{2007}]{Hwang-Noh-2007-Newtonian}
         Hwang J., Noh H., 2007, J. Cosmol. Astropart. Phys., {12}, 003

\bibitem[\protect\citeauthoryear{{Hwang} and {Noh},}{{Hwang} \& {Noh}}{2013}]{Hwang-Noh-2013}
         Hwang J., Noh H., 2013, MNRAS, {433}, 3472

\bibitem[\protect\citeauthoryear{{Jedamzik},}{{Jedamzik}}{2010}]{GW-Jedamzik}
         Jedamzik, K., Lemoine, M., Martin, J., 2010, J. Cosmol. Astropart. Phys., 04, 021

\bibitem[\protect\citeauthoryear{{Jeong} et~al.,}{{Jeong} et~al.}{2011}]{JGNH-2011}
         Jeong D., Gong J., Noh H., Hwang J., 2011, ApJ, {722}, 22

\bibitem[\protect\citeauthoryear{{Jeong} and Schmidt,}{{Jeong} \& {Schmidt}} {2015}]{Jeong-Schmidt-2015}
			Jeong D., Schmidt F., 2015, Class. Quantum Gravity 32, 044001

\bibitem[\protect\citeauthoryear{{Jeong} et al.,}{{Jeong} et al.} {2012}]{Jeong-etal-2012}
			Jeong D., Schmidt F., Hirata, C. M., 2012, Phys. Rev. D, 85, 023504

\bibitem[\protect\citeauthoryear{{Kodama} and {Sasaki},}{{Kodama} \& {Sasaki}}{1984}]{Kodama-Sasaki-1984}
         Kodama H., Sasaki M., 1984, Prog. Theor. Phys. Suppl., 78, 1

\bibitem[\protect\citeauthoryear{{Komatsu} et al.,}{{Komatsu} et al.}{2009}]{wmap5}
Komatsu E., et al., 2009 ApJS, 180, 330

\bibitem[\protect\citeauthoryear{{Lewis} et al.,}{{Lewis} et al.}{2000}]{camb}
Lewis A., Challinor, A., Lasenby, A., 2000, ApJ, 538, 473

\bibitem[\protect\citeauthoryear{{Lifshitz},}{{Lifshitz}}{1946}]{Lifshitz-1946}
         Lifshitz E. M., 1946, J. Phys., {10}, 116

\bibitem[\protect\citeauthoryear{{Mollerach},}{{Mollerach}}{2004}]{GW-Mollerach}
         Mollerach, S., Harari, D., Matarrese, S., 2004, Phys. Rev. D, {69}, 063002

\bibitem[\protect\citeauthoryear{{Noh},}{{Noh}}{2014}]{Noh-2014}
         Noh H., 2014, J. Cosmol. Astropart. Phys., {07}, 037

\bibitem[\protect\citeauthoryear{{Noh} and {Hwang},}{{Noh} \& {Hwang}}{2004}]{Noh-Hwang-2004}
         Noh H., Hwang J., 2004, Phys. Rev. D, {69}, 104011

\bibitem[\protect\citeauthoryear{{Peebles},}{{Peebles}}{1980}]{Peebles-1980}
         Peebles P. J. E., 1980, The Large-scale Structure of the Universe. Princeton Univ. Press, Princeton, NJ

\bibitem[\protect\citeauthoryear{{Sarkar},}{{Sarkar}}{2008}]{GW-Sarkar}
         Sarkar, D., Serra, P., Cooray, A., Ichiki K., Baumann, D., 2008, Phys. Rev. D, {77} 103515

\bibitem[\protect\citeauthoryear{{Vishniac},}{{Vishniac}}{1983}]{Vishniac-1983}
         Vishniac E. T., 1983, MNRAS, {203}, 345

\bibitem[\protect\citeauthoryear{{Weinberg},}{{Weinberg}}{1987}]{Weinberg-1987}
         Weinberg S., 1987, Phys. Rev. Lett., {59}, 2607

\bibitem[\protect\citeauthoryear{{Yoo},}{{Yoo}} {2010}]{Yoo-2010}
			Yoo J., 2010, Phys. Rev. D, 84, 063505

\bibitem[\protect\citeauthoryear{{Yoo},}{{Yoo}} {2014}]{Yoo-2014}
			Yoo J., 2014, Class. Quant. Gravity, 31, 234001

\bibitem[\protect\citeauthoryear{{Yoo} et al.,}{{Yoo} et al.} {2009}]{Yoo-etal-2009}
			Yoo J., Fitzpatrick A. L., Zaldarriaga M., 2009, Phys. Rev. D, 80, 083514

\bibitem[\protect\citeauthoryear{{Yoo} and {Zaldarriaga},}{{Yoo} \& {Zaldarriaga}}{2014}]{Yoo-Zaldarriaga-2014}
			Yoo J., Zaldarriaga, M., 2014, Phys. Rev. D, 90, 023513

\bibitem[\protect\citeauthoryear{{York},}{{York}}{1973}]{York-1973}
         York J. W., 1973, J. Math. Phys., 14, 456

\end{thebibliography}
\end{document}